\documentclass{article}
\usepackage[T1]{fontenc} 
\usepackage[utf8]{inputenc} 
\usepackage{ismir,amsmath,cite,url}
\usepackage{graphicx}
\usepackage[bookmarks=false]{hyperref}
\hypersetup{
  colorlinks=false,      
  pdfborder={0 0 0},     
}

\usepackage{color}

\usepackage[firstpage]{draftwatermark}
\definecolor{lightgray}{rgb}{0.9,0.9,0.9}
\definecolor{darkgray}{rgb}{0.4,0.4,0.4}
\SetWatermarkFontSize{12pt}
\SetWatermarkScale{1.1}
\SetWatermarkAngle{90}
\SetWatermarkHorCenter{202mm}
\SetWatermarkVerCenter{170mm}
\SetWatermarkColor{darkgray}
\SetWatermarkText{Late-Breaking / Demo Session Extended Abstract, ISMIR 2024 Conference}

\def\lbd{}	

\title{MusicGen-Chord: Advancing Music Generation through Chord Progressions and Interactive Web-UI}

\multauthor
{Jongmin Jung$^{1,2}$ \hspace{1cm} Andreas Jansson$^2$ \hspace{1cm} Dasaem Jeong$^3$} {
 $^1$Dept. of Artificial Intelligence, $^3$Dept. of Art \& Technology, Sogang University , Seoul, South Korea\\
$^2$Replicate, San Francisco, USA\\
{\tt\small \{jongmin, dasaemj\}@sogang.ac.kr, andreas@replicate.com}
}

\def\authorname{J. Jung, A. Jansson, and D. Jeong}

\begin{document}

\maketitle
\begin{figure}
 \centerline{
 \includegraphics[alt={Image of melody and chord chromagram matrices},width=1.1\columnwidth]{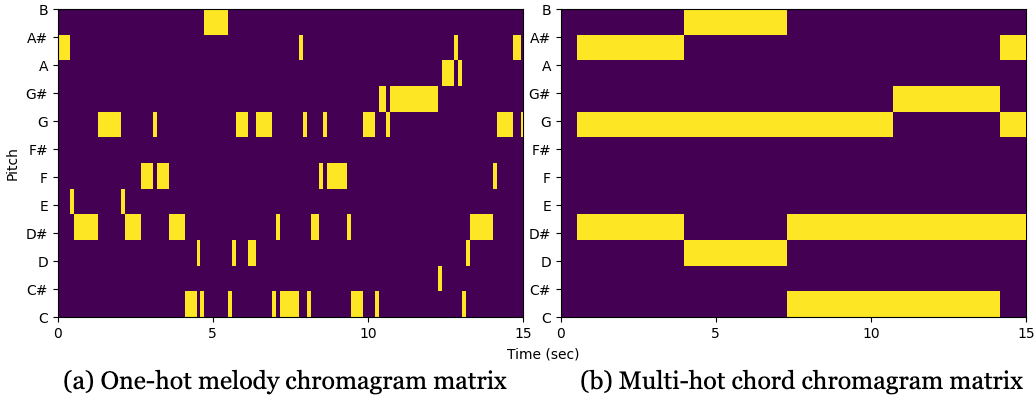}}
 \caption{(a) MusicGen's melodic features in a matrix of one-hot encoded chroma vectors. (b) MusicGen-Chord's chord progression features in a matrix of multi hot encoded chroma vectors. For example, in the chromagram above right there is an E$\flat$ chord (E$\flat$, G, B$\flat$) followed by a G chord (D, G, B), followed by a C minor chord (C, E$\flat$, G), etc.}
 \label{fig:chroma}
\end{figure}

\begin{abstract}
MusicGen is a music generation language model (LM) that can be conditioned on textual descriptions and melodic features. We introduce \textit{MusicGen-Chord}\footnote{\url{https://replicate.com/sakemin/musicgen-chord}}, which extends this capability by incorporating chord progression features. This model modifies one-hot encoded melody chroma vectors into multi-hot encoded chord chroma vectors, enabling the generation of music that reflects both chord progressions and textual descriptions. Furthermore, we developed \textit{MusicGen-Remixer}\footnote{\url{https://replicate.com/sakemin/musicgen-remixer}}, an application utilizing MusicGen-Chord to generate remixes of input music conditioned on textual descriptions. Both models are integrated into Replicate’s web-UI using \texttt{cog}, facilitating broad accessibility and user-friendly controllable interaction for creating and experiencing AI-generated music.
\end{abstract}
\section{Introduction}\label{sec:introduction}
The trend in generative AI emphasizes the controllability of models, allowing users to direct and refine outputs according to their preferences. Notable examples include Stable Diffusion~\cite{stablediffusion,SDXL}, supported by interfaces like \texttt{AUTOMATIC1111’s web-UI}~\cite{AUTOMATIC1111} and \texttt{ComfyUI}~\cite{ComfyUI}, which offer extensive user control over image generation processes. In the realm of music generation, models with enhanced controllability are emerging, offering conditions such as chord\cite{cocomulla,jasco,musicongen}, rhythm\cite{cocomulla,jasco,musicongen}, melody\cite{Musicgen,jasco}, and style based on reference audio \cite{musicgenstyle}. This paper explores the integration of controllability in music generation through the example of MusicGen-Chord.

MusicGen~\cite{Musicgen} is an auto-regressive, Transformer-based music generation model that enables user control through textual descriptions and melodic features. It processes multiple streams of compressed discrete audio representations~\cite{encodec} to generate high-quality, coherent, and stylistically diverse music. MusicGen-Chord extends this model by conditioning on chord progressions instead of melodies. This modification uses a matrix of multi-hot encoded chroma vectors to represent chord progression features. MusicGen-Chord was released in October 2023, and since then, several similar but more advanced studies have been introduced, such as MusiConGen\cite{musicongen}.

To demonstrate the practical benefits of this approach, we developed MusicGen-Remixer, an application based on MusicGen-Chord. This application allows users to upload a music track, provide a textual description prompt, and generate a new background track that is remixed with the input audio. By leveraging Replicate’s web-UI and the \texttt{cog}~\cite{cog} package, MusicGen-Remixer and MusicGen-Chord are made widely accessible on the cloud, promoting user-friendly interaction and broad accessibility for creating and experiencing AI-generated music.
\section{MusicGen-Chord}
MusicGen-Chord extends the original MusicGen model by shifting the conditioning target from melodies to chord progressions. The original MusicGen model uses one-hot encoded chroma vectors as input condition to represent melodies (\figref{fig:chroma}.(a)). In this approach, each vector indicates the presence of a single pitch class at a given time, which is effective for simple melodies but limited in capturing complex harmonic content.

We found that we can tweak this input format to a multi-hot format to represent chord condition (\figref{fig:chroma}.(b)). These multi-hot chroma vectors can encode multiple active pitch classes for each time frame, providing a more comprehensive representation of harmonic structures. This ``trick'' works surprisingly well, enabling MusicGen-Chord to generate chord progressions that align with the style indicated by the prompt using the pretrained MusicGen model weights, \textit{without} requiring any fine-tuning.

The interface for MusicGen-Chord accepts chord progression inputs both as audio and text formats, automatically converting them into multi-hot chroma vectors for the model. This flexibility allows users detailed control over the harmonic structure, enabling a more interactive and customizable music creation experience. For text-based inputs, users can specify chords using a simple format that includes the root and type of each chord, defined by \texttt{ROOT:TYPE}~\cite{harte}. Each chord lasts for a single bar, with the option to add multiple chords in a bar by separating them with commas. For example: \texttt{"G:maj7 D:min7,G:7 C:maj7 F:7 B:min7,Bb:7 A:min7,D:7"}. These text inputs are converted into chroma representations based on the input BPM value. For audio-based inputs, a chord extraction model, BTC~\cite{BTC} is employed to predict symbolic chords along with their timestamps, which are then represented in the chord chroma feature, ensuring effective incorporation of the harmonic content into the music generation process.

\section{MusicGen-Remixer}
MusicGen-Remixer utilizes the features of MusicGen-Chord to enable the creation of remixed music tracks. This application allows users to upload a music track, provide a textual description prompt, and generate a new background track that is remixed with the input audio.

The process involves several sophisticated steps to ensure the generation of coherent and contextually relevant remixes:\footnote{\url{https://github.com/sakemin/musicgen-remixer}}
\begin{enumerate}
    \item \textbf{Input Music Structure Analysis:} Utilizing the \texttt{All-in-One}\cite{taejun2023allinone} framework, the input music’s BPM and downbeats are detected to maintain temporal integrity.
    \item \textbf{Source Separation:} A neural source separation model, Demucs\cite{demucs} is employed to separate vocal tracks from instrumental components, ensuring the original vocal performance is preserved.
    \item \textbf{Chord Progression Feature Extraction:} BTC is used to extract chord progression features from the input audio, guiding the generation of the new background track.
    \item \textbf{Dynamic Time Warping:} Using \texttt{Py-TSMod}\cite{pytsmod}, the timing of the generated track is adjusted to match the downbeats of the input audio, ensuring rhythmic consistency.
    \item \textbf{Mixing:} The aligned background track is mixed with the separated vocal track to produce a cohesive remixed output.
\end{enumerate}
\begin{figure}
 \centerline{
 \includegraphics[alt={Image of melody and chord chromagram matrices},width=1.1\columnwidth]{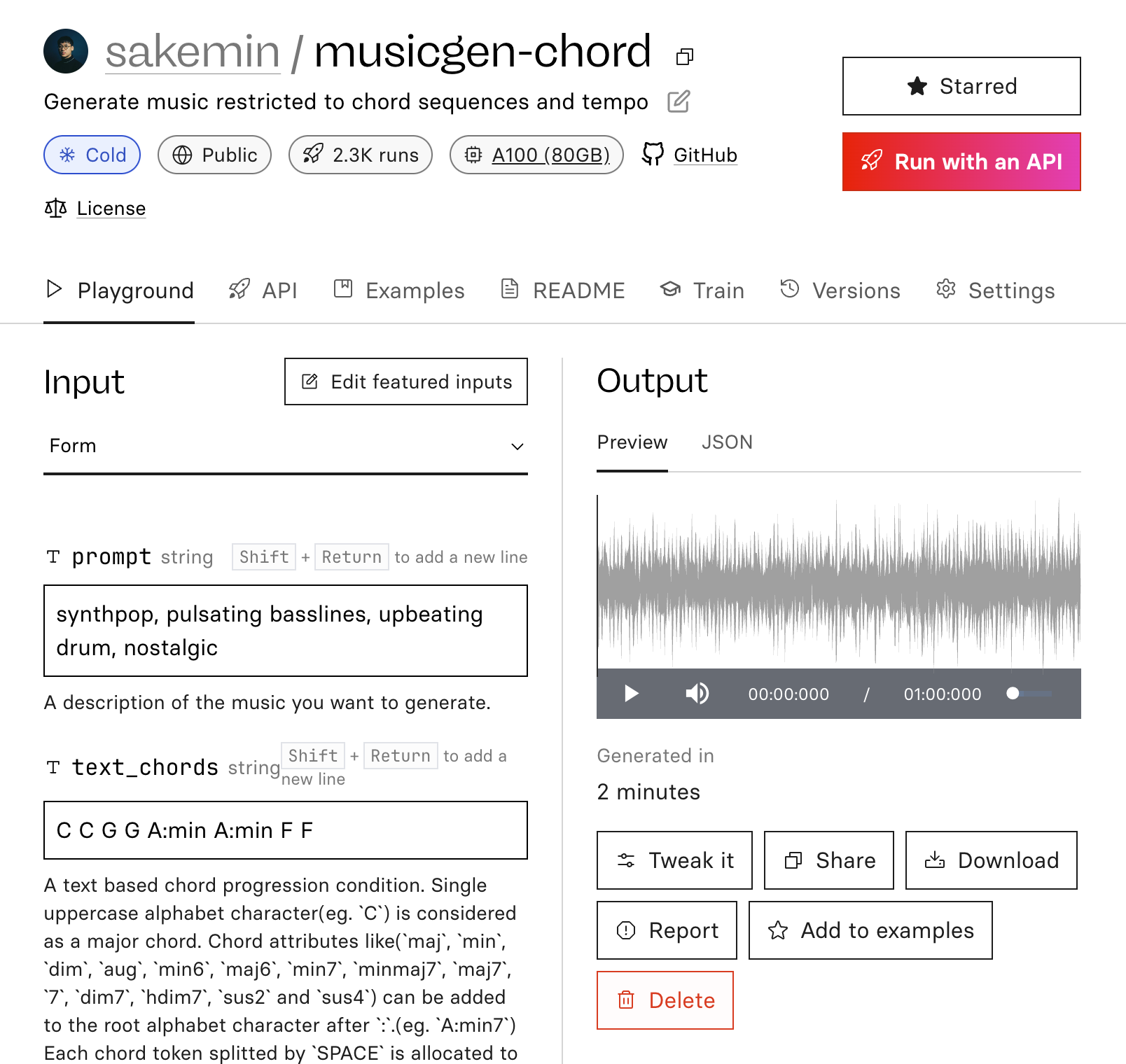}}
 \caption{Replicate's webUI demo of MusicGen-Chord.}
 \label{fig:replicate}
\end{figure}
\section{REPLICATE INTEGRATION}
Replicate’s web-UI, combined with the \texttt{cog} package, provides a seamless and convenient platform for deploying AI models. The \texttt{cog} package can encapsulate AI models with all their dependencies, including Python packages, operating system components, and CUDA versions. This integration ensures that models are portable and easily deployable, facilitating a user-friendly interface for model interaction and management .

MusicGen-Chord and MusicGen-Remixer are integrated with Replicate through the \texttt{cog} package, making them widely accessible on the cloud. Users can easily interact with these models via the web interface, API, or directly using the \texttt{cog}-wrapped repository\footnote{\url{https://github.com/sakemin/cog-musicgen-chord}} on local machines. This setup ensures a straightforward and accessible experience for generating and remixing AI-driven music, enhancing both usability and adaptability.

Replicate offers a practical solution for sharing AI model demonstrations within the MIR community. Researchers and developers can utilize Replicate as an effective tool for presenting and disseminating their work, as demonstrated by the Music Technology Group (MTG)\footnote{\url{https://replicate.com/mtg}} and others who have successfully used this platform. For example, we implemented MusiConGen\cite{musicongen}, a recent MusicGen variant with controllable chord and rhythm features, into a \texttt{cog} wrapped demo\footnote{\url{https://replicate.com/sakemin/musicongen}}.
\bibliography{ISMIRtemplate}

\end{document}